\documentclass[a4paper,twoside]{article}

\usepackage{epsfig}
\usepackage{subcaption}
\usepackage{calc}
\usepackage{amssymb}
\usepackage{amstext}
\usepackage{amsmath}
\usepackage{amsthm}
\usepackage{multicol}
\usepackage{pslatex}
\usepackage{apalike}
\usepackage{algorithm2e}
\usepackage[bottom]{footmisc}

\usepackage[table,xcdraw]{xcolor}
\usepackage{amsfonts}
\usepackage{textcomp}
\usepackage{units}
\usepackage{url}
\usepackage{graphicx}
\def\imagescale{0.99}

\usepackage{SCITEPRESS}     % Please add other packages that you may need BEFORE the SCITEPRESS.sty package.

\begin{document}

\title{Using The Polynomial Particle-in-Cell Method for Liquid-Fabric Interaction}

\author{\authorname{{Robert Dennison}, {Steve Maddock}}
\affiliation{The University of Sheffield}
\email{\{rtdennison1, s.maddock\}@sheffield.ac.uk}
}


\abstract{Liquid-fabric interaction simulations using particle-in-cell (PIC) based models have been used to simulate a wide variety of phenomena and yield impressive visual results. However, these models suffer from numerical damping due to the data interpolation between the particles and grid.
Our paper addresses this by using the polynomial PIC (PolyPIC) model instead of the affine PIC (APIC) model that is used in current state-of-the-art wet cloth models. Theoretically, PolyPIC has lossless energy transfer and so should avoid any problems of undesirable damping and numerical viscosity.
Our results show that PolyPIC does enable more dynamic coupled simulations.
The use of PolyPIC allows for simulations with reduced numerical dissipation and improved resolution of vorticial details over previous work.  For smaller scale simulations, there is minimal impact on computational performance when using PolyPIC instead of APIC. However, as simulations involve a larger number of particles and mesh elements, PolyPIC can require up to a $2.5\times$ as long to generate $4.0s$ of simulation due to a requirement for a decrease in timestep size to remain stable.
}

\onecolumn \maketitle \normalsize \setcounter{footnote}{0} \vfill

\begin{figure*}[htb]
  \includegraphics[width=\imagescale\linewidth]{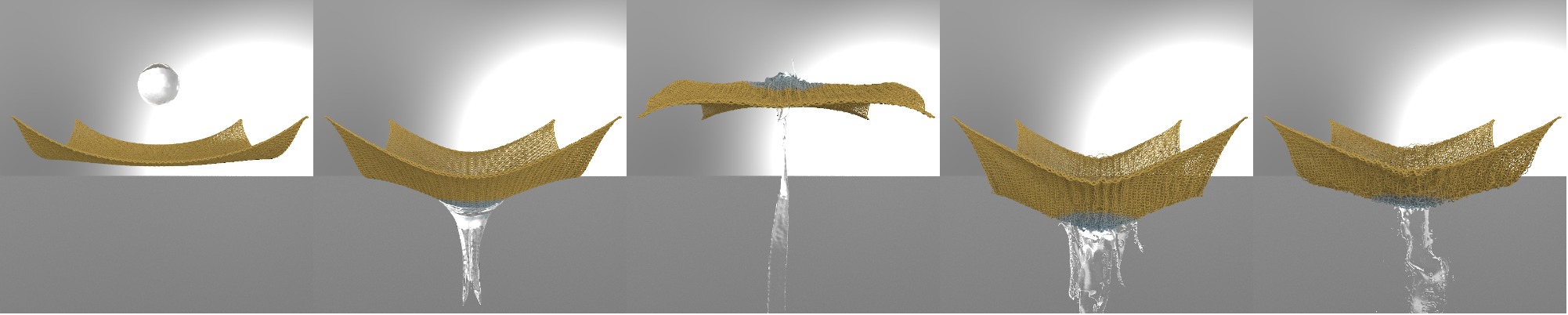}
  \centering
  \caption{Fluid splashing onto a square of yarn fabric using PolyPIC transfers.}
 \label{fig:yarns_medium_render}
\end{figure*}

\section{\uppercase{Introduction}}
\label{sec:introduction}
In physics-based simulations for computer graphics, increasing attention is being focused on simulations involving a combination of two or more physical media.
One area of particular interest is simulating the complex interactions between fluid and porous fabric.
The liquid-fabric interaction model of Fei et al \cite{fei_multi-scale_2018} is widely considered as the state-of-the-art for this kind of coupled simulation. They make use of the affine particle-in-cell (APIC) model for the fluids and the material point method (MPM) using APIC transfers for cloth and yarn objects.

Particle-in-cell (PIC) methods are a class of hybrid Eulerian-Lagrangian simulation methods, designed to benefit from the ease of particle advection found in Lagrangian methods with the simplicity of calculating forces and material properties on a regular Eulerian grid. In each time step, particle information is interpolated to nearby grid nodes where an updated velocity can be calculated. This new velocity can then be interpolated back to the particles which can be advected through the simulation domain.
A known issue with PIC methods is that they suffer from numerical dissipation, as the interpolation stages act as a filter of high frequency and rotational velocities. This means that fluid simulations using PIC methods can seem overly viscous.
APIC was introduced as an improvement to the standard PIC model \cite{jiang_affine_2015} and was developed to reduce the numerical dissipation by considering rotational velocity. Since its introduction, APIC has seen wide adoption by the visual effects industry.

While APIC was developed in an effort to reduce damping of rotational velocities, it has been shown to still be introduce numerical damping.
This makes it more challenging to use real world values for viscosity and elasticity, as the model will introduce viscosity numerically.
The polynomial particle-in-cell model (PolyPIC) was introduced as a generalised extension of APIC to further reduce numerical dissipation \cite{fu_polynomial_2017}, and has been shown to theoretically be able to achieve lossless energy transfers during the interpolation stages.
%{\color{orange} *OLD* By using methods with reduced energy loss, the control over viscosity and elasticity can be improved, making these models more useful.}
By using a method with reduced energy loss, real world empirical data can be used as the model parameters, allowing for easier recreation of real world scenes by artists.

Our paper is the first to use PolyPIC for liquid-fabric interactions.
We improve the work of Fei et al by incorporating polynomial transfers between the particles and grid.
By substituting the affine transfers of APIC for higher order polynomials, the numerical damping of the model can be reduced.
This also requires an alteration of some aspects of the presented PolyPIC method to improve simulation stability.
The results give a comparison of PolyPIC with APIC for coupled simulation scenarios and demonstrate that PolyPIC provides improved preservation of rotational velocity, leading to more dynamic simulations.
The presented scenarios give a comparison of APIC and PolyPIC in a pure fluid simulation to showcase the differences of the two simulation approaches.

The remainder of this paper is structured into five sections. Section \ref{sec:related} describes related work. Section \ref{sec:method} explains the implementation of PolyPIC for the liquid-fabric simulation. The results in section \ref{sec:results} provide a comparison of PolyPIC and APIC, with a discussion in section \ref{sec:discussion}. Section \ref{sec:conclusion} presents conclusions.

%-------------------------------------------------------------------------

\section{\uppercase{Related Work}}\label{sec:related}

Early work on wet cloth models focused on non-porous thin shells interacting with fluids (e.g. \cite{harada_real-time_2007}), with subsequent work involving porous volumes \cite{lenaerts_porous_2008}. The main drawback of this work was that it used a smoothed particle hydrodynamics framework, thus requiring a large number of particles internal to the object's volume and impacting on computational performance. Huber et al \cite{huber_wet_2011} improved the computational performance by using a cellular automata approach which allowed for simple parallelization.
Patkar et al \cite{patkar_wetting_2013} used a mesh-based approach for solid simulation to remove the need for simulating large numbers of solid particles and also incorporated additional fluid effects such as dripping and surface flow.

Further developments came with the application of position based dynamics \cite{muller_position_2007} to the model presented by Lenaerts et al, thus allowing for larger time steps to be used \cite{shao_position-based_2018}.
Another recent development by Fei et al \cite{fei_multi-scale_2018} demonstrates both a porous cloth model using discrete shells \cite{grinspun_discrete_2003} and discrete rods \cite{bergou_discrete_2008} discretized using the material point method (MPM) \cite{sulsky_particle_1994} for cloth and yarn simulations, respectively.
Zheng et al then added further realism to liquid-yarn interactions by modelling the defects in the constituent fibres of a material \cite{zheng_stains_2021}.
However, this work has currently only been used to model static fabric.

Particle-in-cell methods have seen extensive use since first being introduced by Harlow et al \cite{harlow_particle--cell_1962}.
The severe numerical dissipation of this method due to filtering caused by the number of interpolation steps led to the development of the fluid implicit particle (FLIP) method, which bypassed some of the interpolation steps to increase the dynamism of simulations by interpolating only the change in velocity from the grid to the particles rather than interpolating the velocity \cite{brackbill_flip_1986}.
The major drawback of the FLIP method was that reduced dissipation came at the cost of introducing more noise and instability to the simulations.

Jiang et al \cite{jiang_affine_2015} developed the APIC method  as a way of reducing the numerical dissipation of PIC in a stable way by storing an affine transform matrix on each particle, as well as a velocity vector.
This reduced the energy loss of the PIC method and improved the stability of FLIP, and also improved the preservation of angular momentum of both previous iterations of the model.
Recently PolyPIC \cite{fu_polynomial_2017} was developed as a generalized extension of APIC to allow for higher order transfers during the grid/particle transfers, further reducing the numerical dissipation and improving preservation of angular momentum. It achieves this by replacing the affine matrix used to store information about angular momentum with a more general polynomial function.

The popularity of PIC methods led to their application to deformable materials by Sulsky et al to create MPM \cite{sulsky_particle_1994}.
Subsequent PIC developments have also continued to be adapted to elastic solids to continually improve MPM simulations, and MPM has been successfully applied to simulate a wide variety of materials such as snow \cite{stomakhin_material_2013}, viscoelastic solids \cite{fang_silly_2019} and even materials with phase changes such as lava or butter \cite{stomakhin_augmented_2014}.
Despite the improvements of PolyPIC over APIC, PolyPIC has not yet been adapted to coupled simulation scenarios. 

%-------------------------------------------------------------------------

\section{\uppercase{Method}}\label{sec:method}

\begin{figure}[htb]
  \centering
  \includegraphics[width=\imagescale\columnwidth]{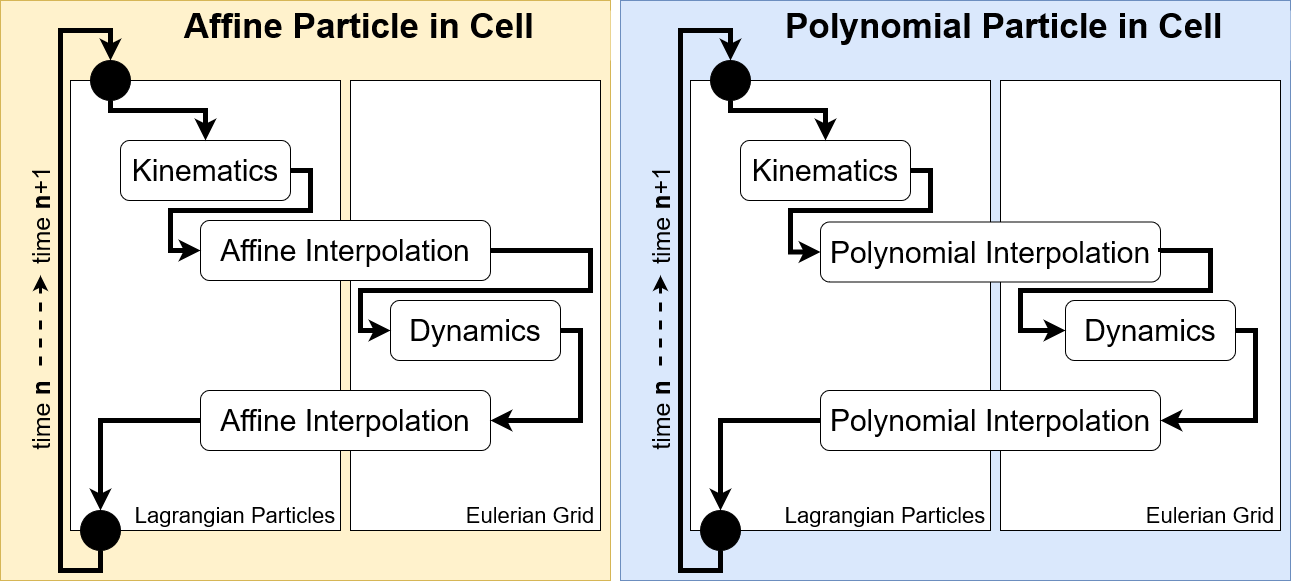}
  \caption{The difference between the APIC and PolyPIC algorithms is the method of transferring data between the particles and the grid.
	PolyPIC uses generalized higher order polynomials whereas APIC uses affine transformation matrices. (Based on Figure 7 in \cite{jiang_affine_2015}.)}
 \label{fig:algorithm_comparison}
\end{figure}

This paper builds on the work of Fei et al by replacing the APIC model with the PolyPIC model.
APIC and PolyPIC differ only in the transfer steps (as shown in Figure \ref{fig:algorithm_comparison}), meaning PolyPIC transfers can be substituted into the model in the place of APIC transfers. Section \ref{sec:pic} presents the standard PIC model, section \ref{sec:apic} describes how APIC builds upon this and PolyPIC is explained in section \ref{sec:polypic}. Finally, section \ref{sec:mixtures} describes the application of PolyPIC to a mixture model.
Fei et al used a staggered marker-and-cell (MAC) grid approach  as it provides more accurate central differences over a standard collocated grid approach, so all work presented here is applied to MAC grids.

% The following subsections give a detailed description of the developments of PIC models.
% Firstly, the standard PIC model is described in subsection \ref{sec:pic}.
% Subsection \ref{sec:apic} describes how APIC builds upon the standard PIC algorithm.
% A detailed explanation of PolyPIC is given in subsection \ref{sec:polypic}.
% Finally, the application of PolyPIC to a mixture model is given in subsection \ref{sec:mixtures}.
% Fei et al used a staggered marker-and-cell (MAC) grid approach  as it provides more accurate central differences over a standard collocated grid approach, so all work presented here is applied to MAC grids.

\subsection{The Particle-in-Cell Method}\label{sec:pic}

\begin{figure}[htb]
  \centering
  \includegraphics[width=\imagescale\columnwidth]{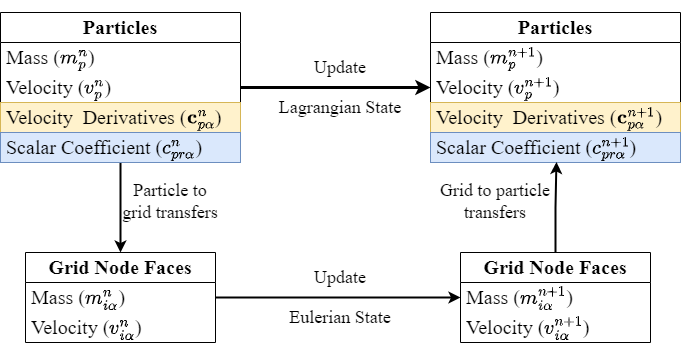}
  \caption{The data stored by particles and grid node faces in PIC methods. White background boxes are consistent between all PIC variants. Velocity derivatives (yellow background) are introduced for APIC and scalar coefficients (blue boxes) are introduced in PolyPIC. (Based on a diagram in \cite{fu_polynomial_2017}.)}
 \label{fig:pic_data}
\end{figure}

The standard PIC model consists of particles which store information about their mass and velocity, which is interpolated to/from the grid node faces to advect the particles around the simulation domain, as shown in Figure \ref{fig:pic_data}. The yellow and blue boxes will be considered in subsections \ref{sec:apic} and \ref{sec:polypic}, respectively.

\begin{table}[htb]
\centering
    {\scriptsize
    \begin{tabular}{|p{3.2em}|p{1.5em}|p{20em}|}
    \hline
    Notation & Type & Meaning \\
    \hline \hline
    $\Delta t$ & s & size of time step \\
    $\Delta x$ & s & size of grid node \\
    $x_p$         & v & position of particle $p$ \\
    $x_{i\alpha}$ & v & position of face $\alpha$ of grid node $i$ \\
    $m_p$         & s & mass of particle $p$ \\
    $v_p$         & v & velocity of particle $p$ \\
    $m_{i\alpha}$ & s & mass of face $\alpha$ of grid node $i$ \\
    $v_{i\alpha}$ & v & velocity of face $\alpha$ of grid node $i$ \\
    $\Tilde{v_{i\alpha}}$ & v & intermediate velocity of face $\alpha$ of grid node $i$ \\
    $w_{ip\alpha}$ & s & contribution of particle $p$ to face $\alpha$ of grid node $i$ \\
    $m_{ip\alpha}$ & s & mass contributed by particle $p$ to face $\alpha$ of grid node $i$ \\
    $(mv)_{ip\alpha}$ & v & momentum contributed by particle $p$ to face $\alpha$ of grid node $i$ \\
    $\textbf{C}_p$ & m & velocity derivatives of particle $p$, as described in \cite{jiang_affine_2015} \\
    $\textbf{c}_{p\alpha}$ & v & velocity derivatives used for MAC grids \\
    $c_{pr\alpha}$ & s & scalar coefficient of scalar mode $r$ for particle $p$ in axis $\alpha$ \\
    $\textbf{I}$ & m & identity matrix \\
    $\textbf{e}_\alpha$ & v & basis vector for axis $\alpha$ \\
    \hline
    \end{tabular}
    }
    \captionof{table}{Notation used throughout this paper. Values are of type scalar s, vector v or matrix m.}
    \label{table:notation}
\end{table}

A PIC simulation consists of a total of $P$ particles and $N_i$ grid nodes, where superscript $n$ is used for a given quantity at the current timestep $n$ (e.g. $v_p^n$ is the velocity of particle $p$ at timestep $n$) and $\alpha$ represents a face of a grid node for each axis $1 \leq \alpha \leq d$, where $d$ is the number of dimensions. The notation used throughout this paper is summarised in Table \ref{table:notation}.

In the standard PIC method, each particle stores its own mass and velocity.
A weight function $N(x)$ is defined to calculate the contribution of a particle to nearby grid cells during the interpolation steps and vice versa.
Each particle within a nearby region of a grid node contributes some of its mass to each face (Equation \ref{eq:pic_momentum_1}).
The mass contribution of each particle can then be multiplied by the particle's velocity to calculate the momentum contribution of each particle (Equation \ref{eq:pic_momentum_2}).
The momentum contribution can then be summed over all particles to calculate the total momentum for the grid node face (Equation \ref{eq:pic_momentum_3}).
The total mass of the grid node face is the mass contributions of all particles (Equation \ref{eq:pic_momentum_4}). 
Finally, the velocity of each grid node face can be calculated by dividing the momentum by the mass (Equation \ref{eq:pic_momentum_5}).

{\footnotesize
\begin{align}
    m_{ip\alpha}^n &= m_p^n w_{ip\alpha} \label{eq:pic_momentum_1} \\ 
    (mv)_{ip\alpha}^n &= m_{ip\alpha}^n v_p \label{eq:pic_momentum_2}\\
    (mv)_{i\alpha}^n &= \sum_{p=0}^P (mv)_{ip\alpha}^n \label{eq:pic_momentum_3}\\
    m_{i\alpha}^n &= \sum_{p=0}^P m_{ip\alpha}^n \label{eq:pic_momentum_4}
\end{align}
}%
{\footnotesize
\begin{align}
    v_{i\alpha}^n &= (mv)_{i\alpha}^n / m_{i\alpha}^n \label{eq:pic_momentum_5}
\end{align}
}%

where $w_{ip\alpha} = N(x_{p\alpha} - x_{i\alpha})$ is the contribution of particle $p$ to face $\alpha$ of grid node $i$.

The mass of a grid node can change throughout the simulation, but the masses of particles are fixed.
This means when interpolating from the grid nodes to the particles, we only update the particles' velocities.
Once the intermediate grid velocities $\Tilde{v_{i\alpha}}^{n+1}$ have been calculated, they are interpolated back to the particles.
To calculate the contribution of grid node momentum to each particle, first the momentum of each grid node must be calculated. This is done by multiplying the mass of each node face by the intermediate velocity of each node face and summing over the number of faces (Equation \ref{eq:pic_transfer_1}).
Then the momentum of each particle can be calculated by summing the contributed momentum of each grid node over the number of grid nodes (Equation \ref{eq:pic_transfer_2}).
Finally, the velocity of each particle can be calculated by dividing the momentum by the particle mass (Equation \ref{eq:pic_transfer_3}).

{\footnotesize
\begin{align}
    (mv)_{ip}^{n+1} &= \sum_{\alpha = 1}^d m_{ip\alpha}^n \Tilde{v_{i\alpha}}^{n+1} \label{eq:pic_transfer_1}\\
    (mv)_{p}^{n+1} &= \sum_{i=0}^{N_i} (mv)_{ip}^{n+1} \label{eq:pic_transfer_2}\\
    v_{p}^{n+1} &= (mv)_{p}^{n+1} / m_p \label{eq:pic_transfer_3}
\end{align}
}%

\subsection{The Affine PIC Method}\label{sec:apic}

In the APIC method, alongside mass and velocity each particle also stores an affine matrix of velocity derivatives to enhance the preservation of rotational velocities (see Figure \ref{fig:pic_data}). Equations \ref{eq:pic_momentum_1} and \ref{eq:pic_momentum_4}, used for transferring mass from the particles to the grid nodes, remain unchanged for APIC. 
The updated momentum transfer takes into account the velocity derivatives to preserve angular momentum (Equation \ref{eq:apic_momentum_1} (Equation 13 in Jiang et al)). The grid momentum can calculated as with PIC using Equation \ref{eq:pic_momentum_3}.

{\footnotesize
\begin{align}
    (mv)_{ip\alpha}^n &= m_{ip\alpha} (\textbf{e}_\alpha v_p + \textbf{c}_{p\alpha}^T(x_p - x_{i\alpha})) \label{eq:apic_momentum_1}
\end{align}
}%

where $m_{ip\alpha}$ is calculated using Equation \ref{eq:pic_momentum_1}.

The momentum transfers from the grid nodes to the particles are the same as those used in the standard PIC method given by Equations \ref{eq:pic_transfer_1} - \ref{eq:pic_transfer_3}.
At this stage the velocity derivatives of a particle, $\textbf{c}_{p\alpha}$, are updated by calculating the gradient of the weights relating that particle to the grid node faces, $\nabla w_{ip\alpha}$,  multiplied by the intermediate velocity, then summing over all grid nodes (Equation \ref{eq:apic_vel_derv} (Equation 14 in Jiang et al)).

{\footnotesize
\begin{align}
    \textbf{c}_{p\alpha}^{n+1} &= \sum_i^{N_i} \nabla w_{ip\alpha} \Tilde{v_{i\alpha}}^{n+1} \label{eq:apic_vel_derv}
\end{align}
}%

\subsection{The Polynomial PIC method}\label{sec:polypic}

PolyPIC replaces the affine matrix of APIC with generalized polynomials which allow for a wider range of local behaviour capture (see Figure \ref{fig:pic_data}).
For PolyPIC the number of modes used is given by $N_r$.
It uses polynomials of the form:

{\footnotesize
\begin{equation}
	s(\textbf{z}) = \prod_{\beta=1}^d z_\beta^{i_\beta}
	\label{eq:polynomial}
\end{equation}
}%

where $z_\beta$ is the $\beta^{th}$ component of $\textbf{z} \in \mathbb{R}^d$ with $i_\beta \in \mathbb{Z}^+$.

Before defining the transfers, we must first define a map of the simulation configuration at time $t^{n+1}$ to $t^n$, denoted as $\xi^{n+1}(x)$. As described by Fu et al, this map can take different forms for constant or affine material motion. Here, the affine map is given by Equation \ref{eq:affine_map} (Equation 10 in Fu et al).

{\footnotesize
\begin{equation}
    \xi^{n+1}(\textbf{x}) = \textbf{x}_p^n + (I + \Delta t C_p^{n+1})^{-1}(\textbf{x} - \textbf{x}_p^{n+1})
    \label{eq:affine_map}
\end{equation}
}%

The method of calculating $\textbf{C}_p^n$ is given in detail in \cite{jiang_affine_2015}. The momentum transfer from the particles to the grid is then calculated by taking the sum of all the scalar modes multiplied by the corresponding scalar coefficients, given by Equation \ref{eq:polypic_momentum} (Equation 11 in Fu et al).
The momentum contribution of each particle to the grid node faces can then be summed over all particles (Equation \ref{eq:pic_momentum_3}). The velocity of each grid node face can then be calculated by dividing the momentum by the mass (Equation \ref{eq:pic_momentum_5}).

{\footnotesize
\begin{align}
    (mv)_{ip\alpha}^n &= m_{ip\alpha}^n \sum_{r=0}^{N_r} s_r (\xi_p^n(x_{i\alpha} - x_p^{n-1}) c_{pr\alpha}^n \label{eq:polypic_momentum}
\end{align}
}%

where $s_r$ is scalar mode $r$.
The coefficients $c_{pr\alpha}^n$ are calculated as a minimisation problem as described by Fu et al.
To efficiently calculate the coefficients, the resultant linear system requires each dimension to be decoupled.
The coefficients for modes $1 \leq r \leq 2^d$ are naturally mass-orthogonal and so solutions can be efficiently found.
However, higher order coefficients $ 2^d < r \leq N_r$ require modification to be orthogonalized.
This is achieved by substituting the quadratic terms, $z_\beta$, in Equation \ref{eq:polynomial} with $g_\beta(z_\beta)$ defined in Equation \ref{eq:orthog}. In contrast to the equation presented in Fu et al, 
the variables in Equation \ref{eq:orthog} are modified to orthogonalize the matrix, which produces more stable simulations.

{\footnotesize
\begin{equation}
    g_\beta(z_\beta) = (w_{ip}^n)^2 - w^n_{ip} \frac{z_\beta(\Delta x^2 - 4z_\beta^2)}{\Delta x^2} - \frac{\Delta x^2}{4} \label{eq:orthog}
\end{equation}
}%

Again, the momentum transfers from the grid to the particles are the same as those used in the standard PIC method given by Equations \ref{eq:pic_transfer_1}, \ref{eq:pic_transfer_2} and \ref{eq:pic_transfer_3}. At this stage, the velocity derivatives $ C_p$ can be calculated using the method described by Jiang et al and the coefficients $c_{pr\alpha}$ can be calculated as described by Fu et al.

\subsection{Liquid-Fabric Interaction}\label{sec:mixtures}

The model presented by Fei et al relies on the use of mixture theory to simulate the interactions between fluid and cloth \cite{nielsen_two-continua_2013}.
Therefore, the momentum transfers need to be altered in order to be applied to mixtures. Let $m_{s,p}^n$ be the mass of a solid particle $p$ at time $n$ and $m_{f,p}^n$ be the mass of a fluid particle $p$ at time $n$. The fluid transfers are then the same as those given in Equations \ref{eq:pic_momentum_3}, \ref{eq:pic_momentum_5} and \ref{eq:polypic_momentum}, except only particles tagged as fluids are considered (Equations \ref{eq:polypic_fluid_1} - \ref{eq:polypic_fluid_3}).
%These fluid momentum transfers are given in Equations \ref{eq:polypic_fluid_1} - \ref{eq:polypic_fluid_3}.

{\footnotesize
\begin{align}
    (m_fv_f)_{ip\alpha}^n &= m_{f,ip\alpha}^n \sum_{r=0}^{N_r} s_r (\xi_p^n(x_{i\alpha} - x_p^{n-1}) c_{pr\alpha}^n \label{eq:polypic_fluid_1}\\
    (m_fv_f)_{i\alpha}^n &= \sum_p (m_fv_f)_{ip\alpha}^n \label{eq:polypic_fluid_2}\\
    v_{f,i\alpha}^n &= (m_fv_f)_{i\alpha}^n / m_{f,i\alpha}^n \label{eq:polypic_fluid_3}
\end{align}
}%

The solid transfers are very similar to the fluid transfers, except that absorbed fluid particles must be considered.
To take account of absorption, we must also consider the mass of absorbed fluid particles when transferring the momentum of solids to the grid.
The total mass of a solid particle is therefore given by summing the mass of the solid particle itself and the mass of the absorbed fluid particle, $(m_{s,ip\alpha}^n + m_{f,ip\alpha}^n)$. These solid-fluid mixture momentum transfers are given in Equations \ref{eq:polypic_solid_1} - \ref{eq:polypic_solid_3}.

{\footnotesize
\begin{align}
    (m_sv_s)_{ip\alpha}^n &= (m_{s,ip\alpha}^n + m_{f,ip\alpha}^n) \sum_{r=0}^{N_r} s_r (\xi_p^n(x_{i\alpha} - x_p^{n-1}) c_{pr\alpha}^n \label{eq:polypic_solid_1}\\
    (m_sv_s)_{i\alpha}^n &= \sum_p (m_sv_s)_{ip\alpha}^n \label{eq:polypic_solid_2}\\
    v_{s,i\alpha}^n &= (m_sv_s)_{i\alpha}^n / (m_{s,ip\alpha}^n + m_{f,ip\alpha}^n) 
    \label{eq:polypic_solid_3}
\end{align}
}%

%-------------------------------------------------------------------------

\section{\uppercase{Results}}\label{sec:results}

\begin{figure}[htb]
  \centering
  \includegraphics[width=\imagescale\linewidth]{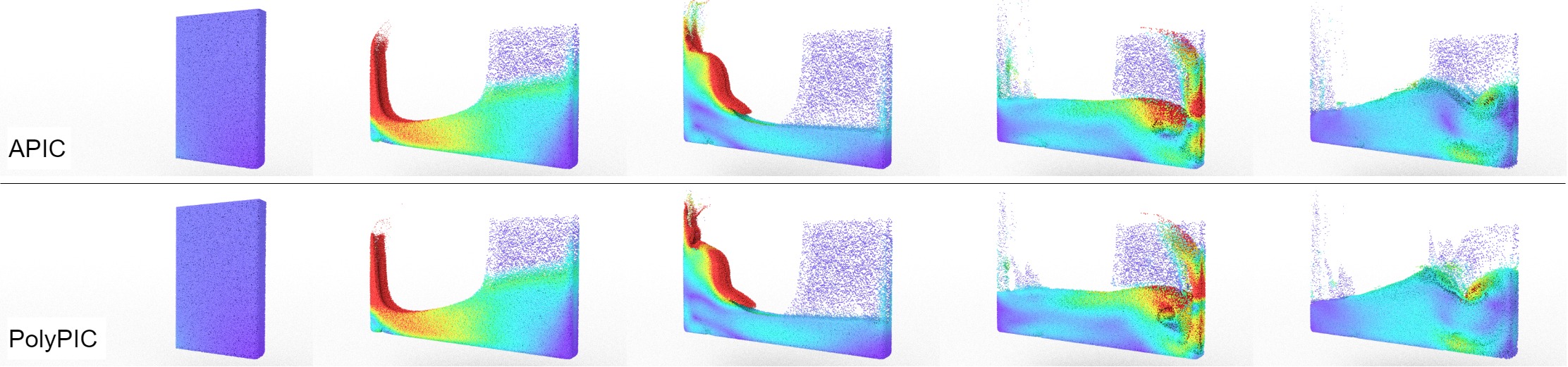}
  \caption{A small dam break scenario using APIC (top) and PolyPIC (bottom) transfers.
	Particle colours indicate velocity (dark blue = low, red = high).
	PolyPIC shows improved resolution of vorticial details.}
 \label{fig:fluid_comparison}
\end{figure}

\begin{figure}[htb]
  \centering
  \includegraphics[width=\imagescale\linewidth]{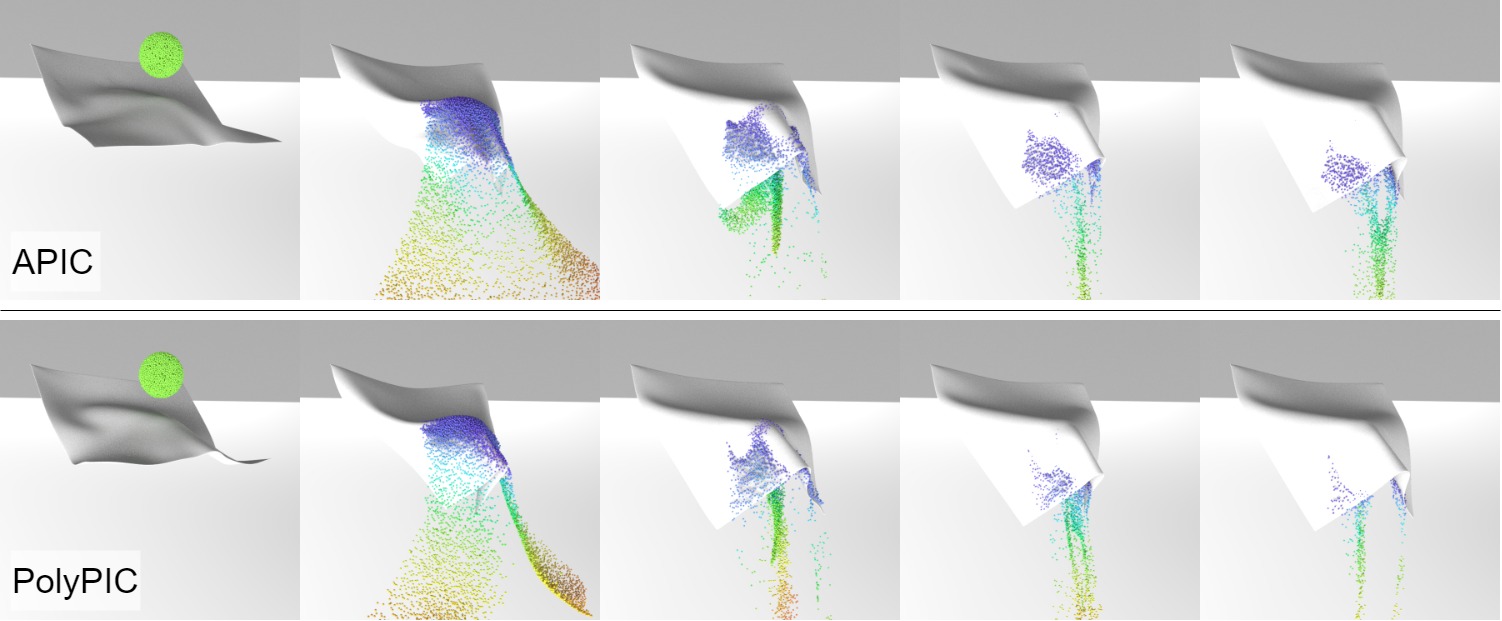}
  \caption{
	Fluid splash onto a small piece of cloth using APIC (top) and PolyPIC (bottom) transfers.
    Particle colours indicate velocity (dark blue = low, red = high).
    The reduced damping means that more fluid splashes off the piece of cloth rather than being absorbed. The fluid that is absorbed exhibits more thin strand behaviour as it drips from the cloth in PolyPIC than APIC (see frame 4).
 }
  \label{fig:cloth_comparison_small}
\end{figure}

\begin{figure}[htb]
    \centering
    \includegraphics[width=\imagescale\linewidth]{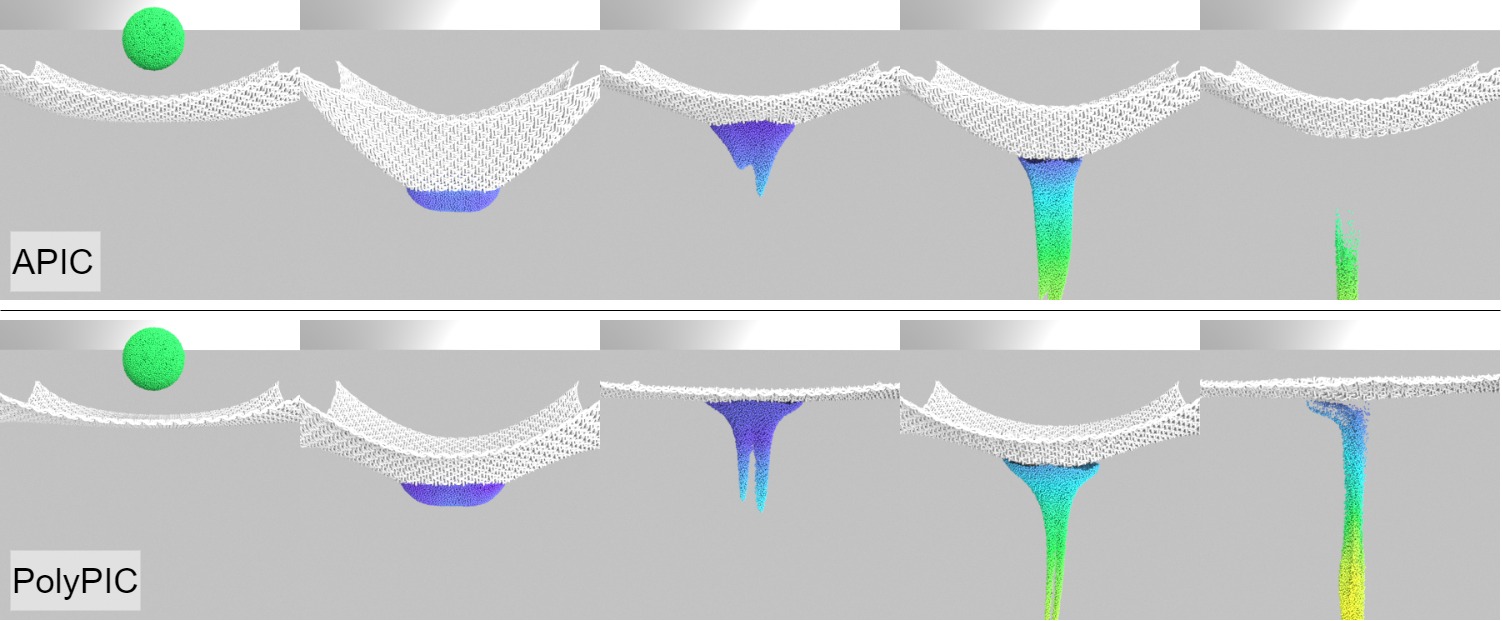}
  \caption{
  Fluid splashing onto a small square of yarn fabric using APIC (top) and PolyPIC (bottom) transfers.
  Particle colours indicate velocity (dark blue = low, red = high).
  The PolyPIC yarns exhibit less sagging than APIC (see frame 1). The reduced damping makes the PolyPIC yarns spring back faster from being stretched (see frames 3-5).
		   }
 \label{fig:yarns_comparison_small}
 \end{figure}

\begin{figure}[htb]
    \centering
    \includegraphics[width=\imagescale\linewidth]{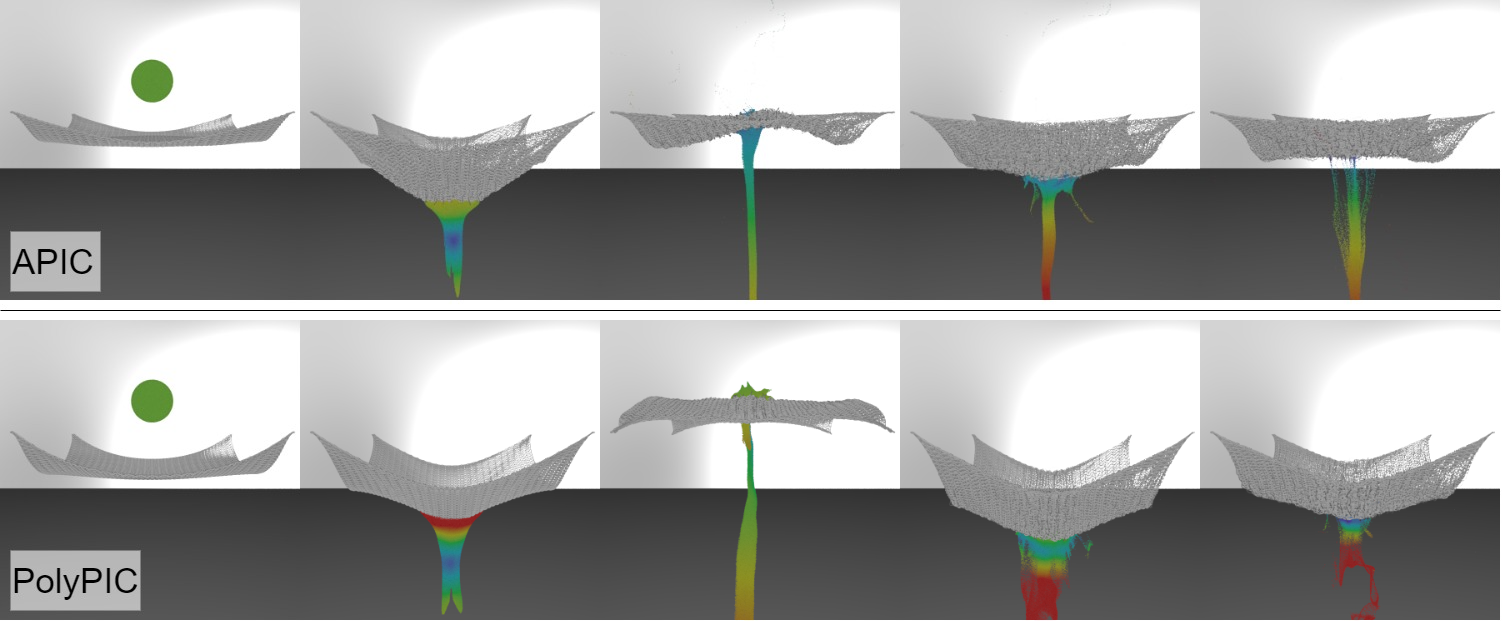}
  \caption{Fluid splashing onto a medium sized square of yarn fabric using APIC (top) and PolyPIC (bottom) transfers. Particle colours indicate velocity (dark blue = low, red = high).
		   PolyPIC causes more energy to be transferred from the falling fluid to the yarn and results in a more dynamic result. 
		   }
 \label{fig:yarns_comparison}

\end{figure}

Four scenarios have been used to test the model that has been developed:

\begin{itemize}
    \item Figure \ref{fig:fluid_comparison} shows a simple dam break scenario to highlight the differences between APIC fluids and PolyPIC fluids.
    \item Figure \ref{fig:cloth_comparison_small} shows a ball of fluid falling onto a small square of cloth draped over a sphere.
    \item Figure \ref{fig:yarns_comparison_small} shows a ball of fluid falling onto a small square of yarn-based fabric draped over a sphere.
    \item Figure \ref{fig:yarns_comparison} is similar to the small yarn scenario but shows a larger volume of liquid falling onto a medium sized square of yarn-based fabric.
\end{itemize}

The latter three scenarios were used by Fei et al, enabling a comparison of the original model using APIC and the new model using PolyPIC.
In the simulation results, particle colours are based on current particle velocity, where dark blue represents a low velocity and red a high velocity.
Videos of each simulation can be found in the supplementary video.

The simple dam break scenario (Figure \ref{fig:fluid_comparison}) highlights the difference between PolyPIC and APIC outside of a coupled simulation context. This shows that using PolyPIC increases the conservation of angular momentum and so improves the resolution of vorticial detail of the fluid. As the final frame shows, PolyPIC particles have higher velocities than in APIC showing that numerical damping is reduced when using PolyPIC.

A ball of fluid falling onto a small square of cloth fabric is shown in Figure \ref{fig:cloth_comparison_small}. This example demonstrates the reduced numerical damping of PolyPIC causes less of the fluid to be absorbed by the cloth, as more splashes off as it is falling. The fluid that is absorbed exhibits more thin strand behaviour as it drips from the cloth in PolyPIC than APIC.

A ball of fluid falling onto a small square of yarn fabric is shown in Figure \ref{fig:yarns_comparison_small}. This scenario shows that the PolyPIC yarns exhibit less sagging than APIC, as they stretch less under their own weight. The reduced damping also makes the PolyPIC yarns spring back from being stretched by the fluid faster.

A breakdown of performance information for each presented example is shown in Table \ref{Table/Results}.
Particles make up the objects that are being simulated.
In the simulation, all particles are flagged as being either fluid or solid particles.
Fluid particles are the particles used to simulate the bulk fluid.
Elements are the mesh elements that make up the fabric (triangles for cloth simulations, rod segments for yarn simulations) and are used for calculating the movement of absorbed fluids.
The `small' simulations use a piece of fabric that is \nicefrac{1}{4} the size of the `medium' simulation (\nicefrac{1}{2} of both the width and height).
Simulations were run using an Intel i9-12900K CPU, and each example was repeated 3 times and the mean performance results are presented. 
The source code can be found at \url{https://github.com/robden820/libwetcloth}. 
For simulations involving PolyPIC, 8 fluid scalar modes and 8 solid scalar modes were used for all examples.

\begin{table*}[htb]
\centering
    \resizebox{0.6\textwidth}{!}{
    \begin{tabular}{|p{10em}|p{3.5em}p{3.5em}p{3.5em}p{3.5em}p{3.5em}p{3.5em}p{3.5em}p{3.5em}|}
    \hline
    Example & {simulation duration (s)} & {timestep (s)} & {s/step (avg)} & total run time (mins) & {peak memory ({\footnotesize{GB}})} & {\#particles (avg)} & {\#fluid particles (avg)} & {\#elements (avg)} \\
    \hline \hline
    \rowcolor[HTML]{cfcfcf} 
     Dam Break Small (APIC)    & 5.0 & 0.0002 & 0.510 & 217 & 1.978 & 310929 & 310929 & 0$^\dagger$ \\
     Dam Break Small (PolyPIC) & 5.0 & 0.0002 & 1.172 & 499 & 2.040 & 310839 & 310839 & 0$^\dagger$ \\
     \hline
    \rowcolor[HTML]{cfcfcf} 
     Splash Cloth Small (APIC)    & 4.0 & 0.0002 & 0.480 & 167 & 0.896 & 12359 & 1360 & 20406 \\
     Splash Cloth Small (PolyPIC) & 4.0 & 0.0001 & 0.428 & 282 & 0.842 & 12228 & 1228 & 20406 \\
     \hline
     \rowcolor[HTML]{cfcfcf} 
     Splash Yarn Small (APIC)    & 4.0 & 0.0002 & 0.454 & 155 & 1.056 & 23074 & 3471 & 19799 \\
     Splash Yarn Small (PolyPIC) & 4.0 & 0.0002 & 0.579 & 197 & 1.109 & 23233 & 3630 & 19799 \\
     \hline
     \rowcolor[HTML]{cfcfcf} 
     Splash Yarn Medium (APIC)    & 4.0 & 0.0002 & 1.897 & 818 & 9.300 & 119520 & 40320 & 79596 \\
     Splash Yarn Medium (PolyPIC) & 4.0 & 0.00005 & 1.460 & 1958 & 5.764 & 116416 & 37213 & 79596 \\
    \hline
    \end{tabular}
    }
    \caption{Timing and storage data for APIC and PolyPIC for three example scenarios. $^\dagger$ The dam break simulations involved no fabric, so required 0 mesh/rod elements.}
    \label{Table/Results}
\end{table*}

As can be seen in Table \ref{Table/Results}, as the number of particles increases, the impact of the additional complexity of the PolyPIC model increases.
The dam break scenarios use the largest number of particles to clearly demonstrate the difference in fluid behaviours when using PolyPIC, but also to highlight the additional computational impact of PolyPIC over APIC.
The small cloth splash scenario has the smallest number of simulated particles and elements and shows the smallest difference in seconds per step and peak memory usage between APIC and PolyPIC of all the examples. 
Some of the simulations require use of a smaller timestep value due to stability issues (see Section \ref{sec:discussion}).

\begin{figure}
    \centering
    \includegraphics[width=0.94\linewidth]{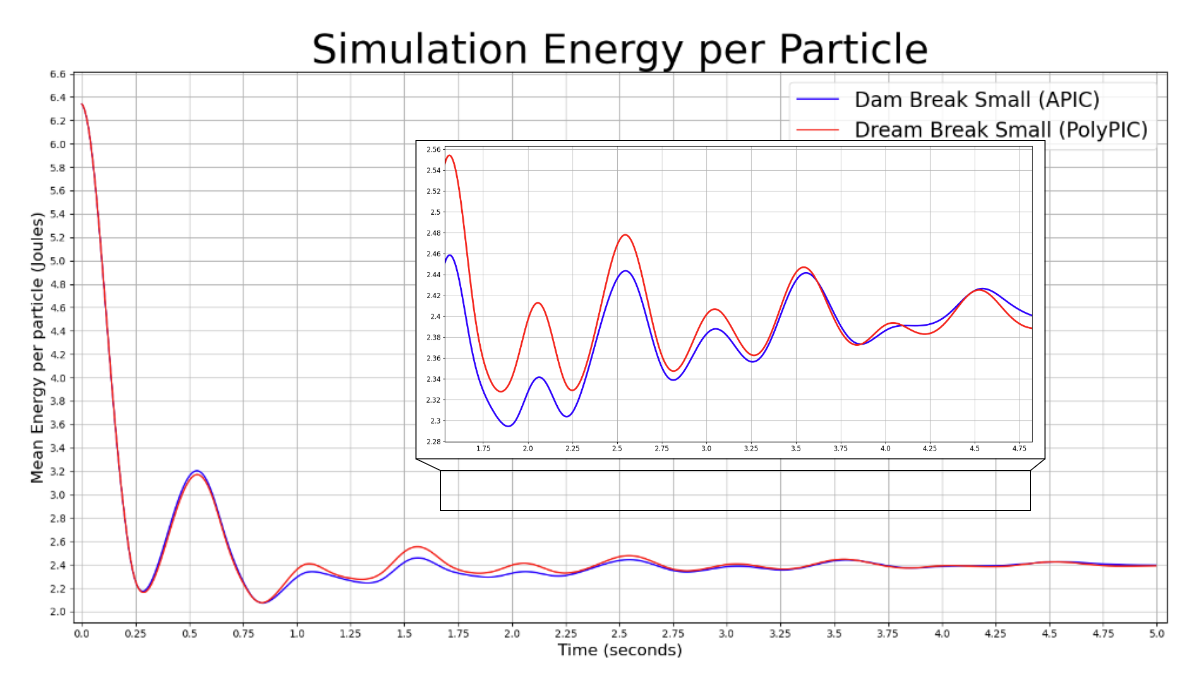}
    
  \caption{Mean particle energy for the dam break scenario, calculated as the sum of kinetic energy and gravitational potential energy. PolyPIC improves the energy preservation over APIC, but the effect per particle is minor.
           }
 \label{fig:energy_dambreak}

\end{figure}

\begin{figure}
    \centering
    \includegraphics[width=0.94\linewidth]{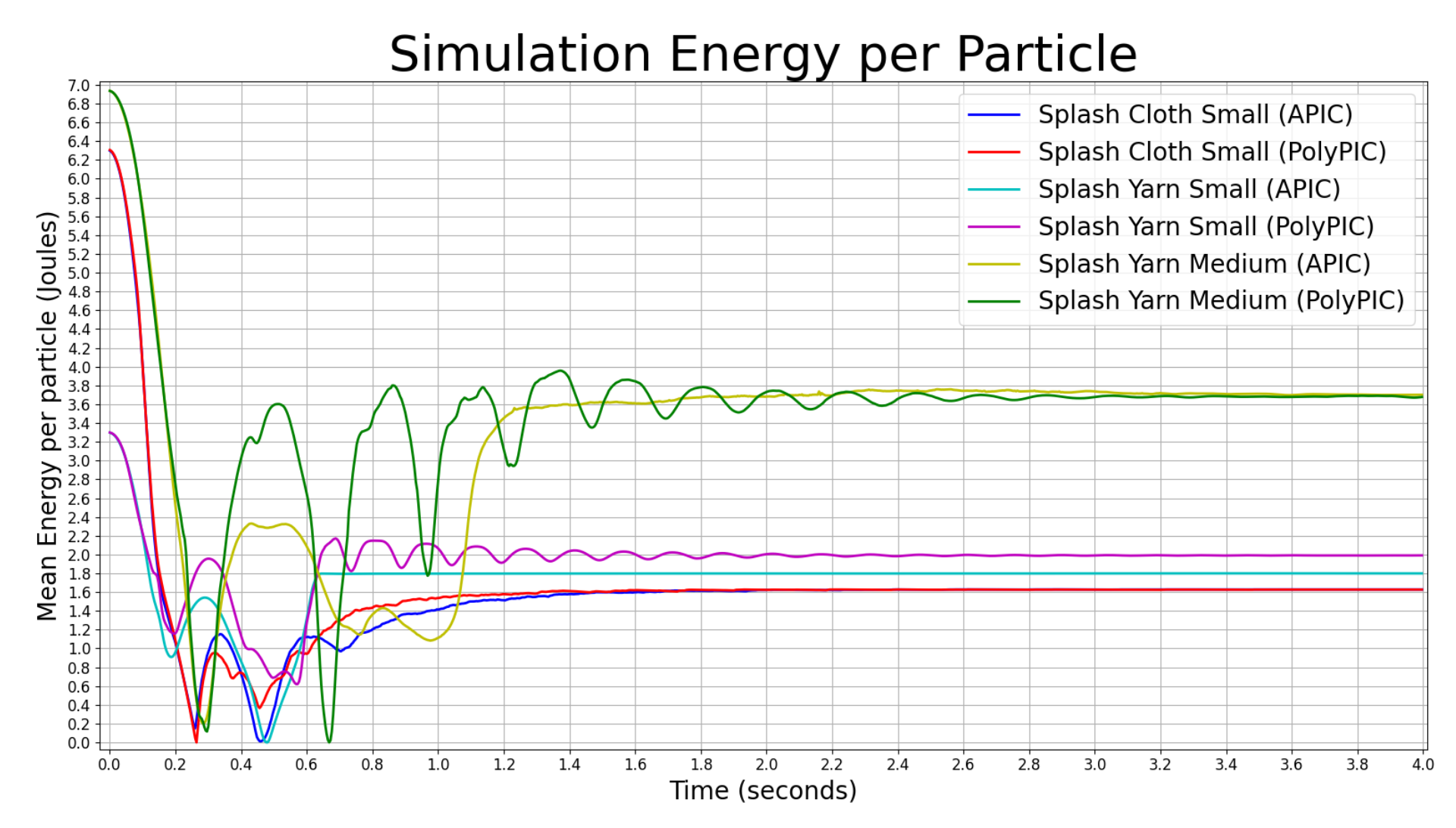}
   \caption{Mean particle energy for the interaction scenarios, calculated as the sum of kinetic energy and gravitational potential energy. Change in energy preservation in PolyPIC and APIC is minimal, although the energy of PolyPIC oscillates for a longer period before reaching a constant value.
  Due to differences in the simulation setup (e.g. fabric at different world heights, rigid sphere in cloth scenario), each simulation stabilises around a different final mean energy.
           }
 \label{fig:energy_all}

\end{figure}

Additionally, an energy plot is presented for the dam break scenario in Figure \ref{fig:energy_dambreak} and for the interactions scenarios in Figure \ref{fig:energy_all} (due to variation in the simulation setup, each simulation stabilises around a different final mean energy).
The plots show the mean energy per particle for the presented scenarios, calculated as the sum of the kinetic energy and gravitational potential energy.
Figure \ref{fig:energy_dambreak} shows that for a simulation involving only fluid particles, the improvement in energy preservation gained by using PolyPIC in the place of APIC is minimal, however, as shown in Figure \ref{fig:fluid_comparison} there is still a difference in the simulations visual output achieved using PolyPIC.
Figure \ref{fig:energy_all} shows a similar trend for the interaction scenarios, and PolyPIC has a small impact in the long term energy preservation of the simulation.
However, simulations involving PolyPIC require more time to reach a stable energy level.
The increased oscillations of PolyPIC before reaching this stable energy level indicate that PolyPIC suffers less from numerical damping than APIC, leading to more dynamic simulations.

%-------------------------------------------------------------------------

\section{\uppercase{Discussion}}\label{sec:discussion}

As shown in the comparison scenarios, visually, PolyPIC provides a notable increase in simulation dynamics. 
Whilst theoretically PolyPIC has been shown to be lossless when transferring velocity between the particles and the grid (shown in the supplementary material of Fu et al), this can't be achieved in practice. For fluids, using polynomial modes involving multi-quadratic terms (i.e. $N_r > 2^d$) causes the simulation to become unstable, even when a very small time step is used.
For solids in a non-coupled scenario, the maximum number of modes can be used without effecting the overall stability of the simulation.
However, when simulating coupled interaction, stability issues become increasingly severe as higher order modes are introduced.

As can be seen in Table \ref{Table/Results}, when comparing the dam break scenarios, an increase in the time required for each simulation step can be seen when using PolyPIC rather than APIC.
In the interaction scenarios, the time per simulation generally decreases, but the average number of fluid particles in the simulation also decreases.
The reduced numerical damping of PolyPIC causes the fluid to splash off the fabric rather than being absorbed, so the fluid particles leave the domain of the simulation faster, and the particles are deleted from the simulation.
This reduces the number of fluid particles to be simulated in later timesteps, allowing each step to be simulated faster.

The fabric-interaction simulations for APIC were both able to be simulated using a timestep of $\Delta t = 2e^{-4}s$, but using this value for the PolyPIC scenarios caused the simulation to become unstable. Whilst the same timestep enabled stable simulation of the small yarn PolyPIC scenario, we found using a timestep of $\Delta t = 1e^{-4}s$ and $\Delta t = 5e^{-5}s$ allowed the small cloth and medium yarn PolyPIC simulation to remain stable for their duration, respectively.
This means that while in general there was only a small change in the time required for each simulation step, the need for a smaller timestep for PolyPIC resulted in an overall increase in the time taken to run the simulation, as shown in Table \ref{Table/Results}.

Possible future work would be to experiment with more stable simulation frameworks such as position-based dynamics \cite{muller_position_2007} which has been shown to reduce the need for small timesteps.
Also, in the yarn examples, the individual yarns quickly become tangled after collision with the fluid. Using a more robust collision handling technique such as incremental potential contact \cite{li_incremental_2020} could improve the stability of the simulations.

%-------------------------------------------------------------------------

\section{\uppercase{Conclusions}}\label{sec:conclusion}

The work presented by Fei et al is widely considered to be the state-of-the-art for liquid-fabric interaction simulations. 
This paper has replaced APIC with PolyPIC in their model and performed a comparison of the PolyPIC model with the APIC model in the context of liquid-fabric interactions. 
Using PolyPIC in place of APIC for liquid-fabric interaction improves the dynamism of simulations, increasing energy transfer between the fluid and cloth/yarn and increasing the resolution of vorticial details and small scale splashes.
However, PolyPIC has a higher computation cost over APIC and the reduced numerical damping of PolyPIC also caused stability issues requiring the use of a smaller time step.
This requirement for smaller timesteps highlights the need for a greater consideration of techniques to improve simulation stability and inter-yarn collision detection. Despite this limitation, this paper demonstrates the potential of PolyPIC as a method of improving liquid-fabric interaction simulations based on PIC methods.

\section*{\uppercase{Acknowledgements}}

This research is supported by a Frank Greaves Simpson Scholarship from the University of Sheffield.

\bibliographystyle{apalike}
{\small
\bibliography{Grapp}}

\end{document}